\documentstyle[preprint,revtex]{aps}
\def\al{$\alpha'$}
\begin{document}
\draft
\preprint{UAHEP-936}
\begin{title}
Classical Stringy Black Holes Modify the Thermal Spectrum
\end{title}
\author{B. Harms and Y. Leblanc}
\begin{instit}
Department of Physics and Astronomy, The University of
Alabama,\\
Tuscaloosa, AL 35487-0324
\end{instit}
\begin{abstract}
Non-local ($\alpha'$) corrections to Schwarzschild black
holes are shown to invalidate the thermodynamical
interpretation of black holes.  In particular, the canonical
and Bekenstein-Hawking temperatures are not equal.  The
particle number density of fields quantized in the
($\alpha'$ modified) black hole background is no longer
thermal.  In the non-perturbative region $\alpha' \to
\infty$
(or $M\to 0$), an analytic continuation to the number
density is shown to vanish
exponentially.
\end{abstract}

\narrowtext

In a previous paper \cite{hl1}, we calculated the stringy
($\alpha'$) classical corrections to the 4-dimensional
Scwarzschild black hole metric and studied their effects on
Hawking's black hole thermodynamics.  It was found\cite{hl1}
that both the area law and the first law of black
hole thermodynamics are invalidated by the $\alpha'$-
perturbation corrections.  The breakdown of the first law
meant the inequality of the canonical and the
Bekenstein-Hawking temperature \cite{bek,hawk}.

In this letter we make use of the above findings to provide
a simple calculation of the quantum particle number density
in an ($\alpha'$-) modified black hole background.  The
so-called back-reaction (quantum gravity) effects are
neglected
here.  We shall find that the breakdown of the first law
implies that the particle spectrum is no longer thermal and
that, on the basis of a reasonable speculation on the strong
(non-perturbative) $\alpha'$-limit
($\alpha'\to \infty$) or small black hole mass ($M\to 0$),
the number density vanishes at least exponentially.  This
result is in agreement with the identification of extreme
neutral black holes as massless elementary particles
\cite{hl2,hl3,hl4,hl5}, contrary to the usual thermal
interpretation (with infinite temperature).

The starting point of our exposition is the well known
$\alpha'$-modified Einstein field equations \cite{green},
\begin{eqnarray}
R_{\mu\nu} + {\alpha'\over{2}} R_{\mu\kappa\lambda\tau}
R_{\nu}^{\kappa\lambda\tau} + O(\alpha'^2) = 0 \; ,
\end{eqnarray}
of a 26-dimensional bosonic string theory.  These equations
are obtained from the requirement of the vanishing of the
tree-level dilaton tadpole in a curved background
\cite{nishi}.
As before \cite{hl1} we assume that all but four spacetime
dimensions have been compactified with the internal space
volume taken equal to unity for simplicity.

Applying perturbation theory about the small Regge slope
\al (\al $= {1\over{2\pi T}}$ where T is the string
tension), the unperturbed metric chosen to be the
4-dimensional Schwarzschild metric, Eq(1), yields the
following solution for the modified black hole spacetime
\cite{hl1}
\begin{eqnarray}
ds^2 = -\lambda^2(r) dt^2 + \lambda^{-2}(r) dr^2 +
r^2(d\theta^2 + \sin^2\theta\; d\phi^2)\; ,
\end{eqnarray}
where,
\begin{eqnarray}
\lambda^2(r) = 1 - {2M\over{r}} -{2M^2\alpha'\over{r}}\bigl(
{1\over{r_+^3}} - {1\over{r^3}}\bigr) + O(\alpha'^2) \; ,
\end{eqnarray}
in which $r_+ = 2M$ is the black hole horizon.

An interesting result of our calculations is that, at least
perturbatively, the non-local \al - effects do not alter the
spacetime topology and the horizon remains unshifted.

According to Gibbons and Hawking \cite{gibb}, black hole
themodynamics can be studied by analytically continuing the
Schwarzschild metric to the Euclidean domain, the period of
the compact Euclidean time representing the inverse
(Bekenstein - Hawking) temperature of black hole systems now
endowed with thermal attributes.  In particular, the
Euclidean action evaluated at the black hole solution has
received an interpretation in terms of thermodynamical
entropy, a result justified by the following WKB expression
for the black hole canonical partition function,
\begin{eqnarray}
Z(\beta_H) \sim e^{-S_E/\hbar} \; .
\end{eqnarray}

In a series of recent papers however \cite{hl2,hl3,hl4,hl5},
we have shown the inconsistency of the thermal (canonical)
ensemble for black holes and have pointed out the
inequivalences between the canonical and microcanonical
ensembles (as reflected by a negative microcanonical
specific heat).  In particular, the equilibrium state was
shown to be very inhomogeneous, and thus very far from a
thermal state \cite{hl2,hl3,hl4,hl5}.

Nevertheless, we wish here to play the devil's advocate and
analyze the effects of stringy non-locality from the
viewpoint of the thermodynamical interpretation.

In Euclidean spacetime the important requirement of the
vanishing of the conical (deficit angle) singularity yields
directly the period (inverse Hawking temperature) of the
compact Euclidean time dimension \cite{hl1},
\begin{eqnarray}
\beta_H = 2\pi\bigl[ \lambda (r){d\over{dr}}\lambda
(r)\bigr]_{r=r_+}^{-1} \simeq 8\pi M\bigl(1 + {3\alpha'
\over{8M^2}} + O(\alpha'^2)\bigr)\; .
\end{eqnarray}
The entropy $S_E$ was calculated in Ref.\cite{hl1}.  To
order \al$^2$ we found,
\begin{eqnarray}
S_E &=& \beta_H \bigl({M\over{2}} + {\alpha'\over{8M^2}} +
O(\alpha'^2)\bigr)\nonumber \\
&=& 4\pi M^2\bigl( 1+{5\alpha'^2\over{8M^2}} +
O(\alpha'^2)
\bigr)  \\
&=& {A\over{4}} \Bigl( 1 + {5\alpha'\over{8M^2}} +
O(\alpha'^2)\Bigr)\; , \nonumber
\end{eqnarray}
in which we used Eq.(5) and where $A = 4\pi r_+^2$, i.e. the
area of the horizon.

The canonical temperature of the gas is given by the usual
thermodynamical relation \cite{hl1}
\begin{eqnarray}
\beta = {dS_E\over{dM}} \simeq 8\pi M \bigl( 1 +
O(\alpha'^2)\bigr)\; .
\end{eqnarray}
Two important implications follow from these results.  First
Eq.(6) clearly shows a modification of the area law of black
hole thermodynamics.  Second, and perhaps more importantly,
the Bekenstein-Hawking temperature, Eq.(5), is not equal to
the above canonical temperature of the supposedly thermal
equilibrium state of the gas,
\begin{eqnarray}
\beta \neq \beta_H \; .
\end{eqnarray}
This is an explicit breakdown of the first law of black hole
thermodynamics.

Canonical quantization of, say, a massless scalar field in
the (\al - modified) classical black hole background (one
neglects the back reaction here) should, pending accuracy of
the thermodynamical description, yield a Planckian
distribution for the particle number density at the
equilibrium canonical temperature $\beta$.  However, because
the \al - effects do not shift the horizon or alter the
topology (perturbatively), the scalar particle Green's
functions turn out to be periodic (in the sense of the Kubo-
Martin-Schwinger condition) with period $\beta_H$.  These
are temperature Green's functions and yield a Planckian
distribution for the particle number density with
temperature $\beta_H^{-1}$.  Therefore,
\begin{eqnarray}
n(\omega)& = &{1\over{e^{\beta_H\omega} - 1}} \nonumber \\
&\simeq& {1\over{e^{\beta\omega} - 1}} - \delta\beta {\omega
e^{\beta\omega}\over{(e^{\beta\omega} - 1)^2}} +
O(\delta\beta^2) \; ,
\end{eqnarray}
where,
\begin{eqnarray}
\delta\beta \equiv \beta_H - \beta \simeq {3\pi\alpha'
\over{M}}
+ O(\alpha'^2) \; .
\end{eqnarray}

Eq.(9) shows that the stringy effects modify the thermal
distribution.  Quantum gravity (back reaction) effects are
widely believed to modify the thermal spectrum even more
drastically.  One can imagine loop effects effectively
creating a cosmological constant term in the action.  Such a
term would completely change the topology of the black hole
spacetime and consequently the particle number density will
very likely not be Planckian.  Black holes with a
cosmological constant usually have two horizons and a metric
described by the de Sitter-Schwarzschild solution.

Eq.(10) shows that small (light) black holes belong to the
non-perturbative domain (\al$ \to \infty$).  From the
viewpoint of string theory, a duality transformation from
strong to weak string tension may prove useful in
the investigation of black holes.  Here, however, we can
only speculate on the behavior of the theory of strong \al.
As noticed in our previous work \cite{hl1} and as is
apparent from Eq.(3), higher order \al - corrections
actually worsen the $r \to 0$ (ultraviolet) behavior of the
black hole spacetime.  A conjecture called the Limiting
Curvature Hypothesis (LCH)\cite{brand} however requires that
the full
gravity theory be singularity free.  This would imply that,
although each individual term in the \al - expansion is
singular, the series should be finite.  The above hypothesis
is based on the seemingly sensible physical requirement of
geodesic completeness.

An analytic continuation to strong \al (\al$ \to \infty$) or
equivalently small black holes ($M \to 0$) was suggested in
Ref.\cite{hl1}.  In this continuation Eq.(3) is the small
\al-expansion of the expression
\begin{eqnarray}
\lambda^2(r) = 1 - {2M\over{r}} e^{\alpha' M({1\over{r_+^3}}
-
{1\over{r^3}})} \; .
\end{eqnarray}

The above form effectively regularizes the black hole
singularity at the origin as gravity is shown to vanish
there (asymptotic freedom).  An \al - modified graviton
propagator can be derived from such a spacetime and was
shown to yield a UV-finite quantum theory of gravity
\cite{hl1}.

In the thermodynamical interpretation, it is interesting to
speculate that Eq.(5) for the
Hawking temperature could be analytically continued to the
small black hole domain as follows\cite{hl1},
\begin{eqnarray}
\beta_H = 8\pi M e^{3\alpha'/8M^2} \; .
\end{eqnarray}
Contrary to the usual case where \al - effects are
neglected, we now find that extreme black holes should have
zero
Hawking temperature.  Notice that Eq.(12) does not follow
from the choice of Eq.(11) for the metric.  Inserting
Eq.(12) into the formula
(Eq.(9)) for the particle number density, one would find the
following asymptotic behavior,
\begin{eqnarray}
n(\omega) \sim \exp (-8\pi M\omega
e^{3\alpha'/8M^2})\;\ \ \
; (M\to 0)\; ,
\end{eqnarray}
a result to be compared with the following behavior for
large black holes,
\begin{eqnarray}
n(\omega) \sim \exp (-8\pi M\omega) \;\ \ \ ; (M\to \infty)
\;.
\end{eqnarray}

The reader will by now surely be aware that
the continuation Eq.(11) of Eq.(3) is not unique.  It maybe
interesting to note that any generalization of Eq.(3)
of the form
\begin{eqnarray}
\lambda^2 = 1 - {2M\over{r}}\sum_{n=0}^{\infty}
C_n\alpha'^{\; n}
M^n\Bigl({1\over{r_+^3}} - {1\over{r^3}}\Bigr)^n \; ,
\end{eqnarray}
with $C_0 = C_1 = 1$ as required by Eq.(3), leads to the
same Hawking temperature,
\begin{eqnarray}
T_H = {1\over{8\pi M}}[1 - {3\alpha'\over{8M^2}}] \; .
\end{eqnarray}
Thus the Hawking temperature goes to zero for $M\to \infty$
and $M^2 = 3\alpha'/8$.  Below the latter value $T_H$
approaches $-\infty$.  Clearly, explicit calculation of the
higher order corrections in \al\ \ are badly needed.

Finally, although the considerations of this letter clearly
show that
the exact thermal nature of the particle spectrum in a black
hole background is spoiled by the non-local stringy effects,
the loss of coherence problem is not resolved by the
inclusion of non-local effects.   The resolution of this
problem must reside in the formulation of a self-consistent
theory of quantum gravity.

\acknowledgments

This research was supported in part by the U.S. Department
of Energy under Grant No. DE-FG05-84ER40141.

\end{document}